\documentclass[vecphys]{svmult}

\usepackage{makeidx}         
\usepackage[bottom]{footmisc}

\usepackage{amssymb}
\newcommand{\h}{\hbar}

\begin{document}

\begin{center}
{\bf \Large The information paradox\footnote{Contribution to ``String Theory and Fundamental Interactions'' --- 
in celebration of Gabriele Veneziano 65th birthday --- eds. M. Gasperini and J. Maharana,  Springer Verlag, Heidelberg, 2007.}}
\titlerunning{The information paradox}

\vskip 50pt
{\bf \large Daniele Amati}

\vskip 15pt
International School for Advanced Studies (SISSA), Trieste, Italy\\
\tt{amati@sissa.it}

\end{center}


\setcounter{footnote}{1}
\begin{flushright}

Los muertos que vos matasteis gozan de buena salud\footnote{Jose'  Zorrilla, ``Don Juan Tenorio'' (1844).} 
\end{flushright}

\section*{Abstract}
The incompatibility between gravity and quantum 
coherence represented by black holes  should be solved by a 
consistent quantum theory that contains gravity as superstring 
theory. Despite many encouraging results in that sense, I question 
here the general feeling of a na\"ive resolution of the paradox. 
And indicate non trivial physical possibilities towards its solution 
that are suggested by string theory and may be further investigated 
in its context.

\section{Introduction}
\label{sec:1}
The fact that black holes represent an apparent contradiction between
 gravity and quantum mechanics is a too well known problem to need 
exhaustive recall. The best way to visualize it is to consider 
together the formation and evaporation processes.  We may envisage a 
b.h. to be formed by a pure quantum state prepared in a distant flat 
space (an impinging spherical wave, two or 25 particles colliding at 
high energies and small impact parameter, and so on).  If the 
characteristics of a b.h. --- including its evaporation implied by 
quantum mechanics --- depend only on few basic parameters (M, Q, J) 
as required by general relativity (no hairs), it is clear that  
quantum coherence of the initial state is totally lost in the 
process. The contradiction has resisted efforts to doctored 
modifications (as corrections to the thermal character of Hawking 
evaporation) and brought distinguished scientists to give up either 
quantum mechanics~\cite{1} or the relevance of classical (general 
relativity) solutions in a path integral formulation of the quantum 
theory of gravitation~\cite{2}.

On the other hand, the advent of string (actually superstring) theory
 as a consistent quantum theory that contains gravity gave confidence
 that somehow the paradox should be solved in its framework. Much 
progress has been done in studying b.h. regimes in string theories 
and a remarkable set of coincidences have been revealed. After 
briefly recalling those results, I will argue that the paradox not 
only isn't trivially solved as often claimed, but manifests its full 
vitality in compelling some quite novel possibilities in the 
generalization to the quantum realm of some classical concepts as 
space-time and its geometry, or the influence that quantum effects 
may have on the actual realization of classical geometrical 
configurations as trapped space-time regions. 

\section{String theories and black holes}
String theories contain arbitrarily massive states within regions 
characterized by the string length $l_s$ --- the basic dimensional 
parameter of the theory --- thus states that, classically, would 
represent black holes. The mass beyond which those states should be 
black holes, depends~\cite{3} on the string coupling g, the other 
--- dimensionless --- basic parameter of the theory. Or in different 
words, for every mass (or excitation energy) there is a small 
coupling string regime, and a large coupling b.h. regime.

In the string regime, D-branes in four~\cite{4} and five~\cite{5} 
dimensions with a convenient number of charges have been studied. 
BPS states have been counted as well as nearly BPS states for 
certain regions of moduli space where perturbative computations are 
feasible~\cite{6}. Decay rates have been computed~\cite{7} -- by 
averaging over the many degenerate initial states --- and shown to 
have a typical thermal distribution. The moduli independence of 
these results allows to conjecture~\cite{8} their validity beyond 
the moduli region  where they were computed. And their g 
independence, also suggested by non-renormalization 
arguments~\cite{9}, may imply their possible continuation beyond 
the weak coupling regime.

An independent treatment --- on totally different grounds --- of the 
strong coupling regime substantiates that impression. The large g 
description of the 4 and 5 dimensional systems just described is 
found by solving the 10-d supergravity equations after reduction on 
the same compact manifold used for the D-brane description. The 
solution generates a metric~\cite{10} that depends on parameters 
that are related to the charges through the moduli of the compact 
manifold. The metric shows an event horizon even in the extreme 
limit in which its area gives the Bekenstein-Hawking entropy of 
extremal b.h.. This entropy and the ADM mass coincide with the 
(exponentiated) multiplicity and mass of the BPS states with the 
same charges as computed from the D-branes in the small coupling 
regime. For nearly extremal b.h. the entropy and the evaporation 
spectrum --- obtained by solving wave equations in the corresponding 
metric background --- coincide again~\cite{7} with those computed 
for small g. And, remarkably, even deviations from black body 
spectrum seem to agree~\cite{11}.

The microscopic formulation of the 5-d near extremal b.h. has been 
further studied~\cite{12} in terms of the D1-D5 brane system. The 
AdS/ CFT correspondence was shown to play a role in the matching 
between supergravity results and the microscopic (SCFT) formulation 
of the b.h. thermodynamics and Hawking radiation, the b.h. being defined 
through a density matrix.   

All these  agreements 
among such different computations gave 
confidence to the g continuation of the theory to a strong coupling 
regime where b.h. physics is met. This direct connection between 
the semiclassical black hole picture and a unitary quantum approach, 
has been considered the sign that the information loss due to b.h. 
could be somehow recuperated~\cite{13}. But how this may be achived
is yet far from clear. In the computations just referred it 
appeared clearly that the thermal Hawking radiation was obtained by 
the averaging over the degenerate microstates that are counted by 
the b.h. entropy, while each microstate would have given rise to a 
complex but non thermal radiation with well defined spectra and 
correlations that carry the precise identity of the microstate 
from which they would have been originated.  This is of course a 
basic characteristics of a microstate (a pure quantum state) 
irrespectively of g.  In other words, the black hole microstates are 
not themselves black holes~\cite{14}. And this not only because of 
the absolute specificity of its radiation, but also by not having 
any signal of an event horizon associated with each of them. This 
last fact is of course expected by sheer consistency: if a b.h. 
microstate would be characterized by an event horizon, it would have 
--- itself --- a Bekenstein entropy and thus would not be a pure 
quantum state. The b.h. appears indeed as the macrostate correctly 
defined by a decoherence procedure --- density matrix --- over the 
many non-blackholish microstates of the theory~\cite{15}.

The obvious consequence of the preceding discussion is that a well 
prepared quantum state (a spherical shell impinging from large 
distances, or a two particle scattering at high energy and small 
impact parameter, etc.) is not expected to give rise to a b.h. even 
if the classical conditions for a gravitational collapse are 
apparently satisfied. 

The possibility that microstates do not have a horizon has been more 
recently proposed in a different context~\cite{13}:  for every 
wrapping of a D-1 brane (whose number defines one of the charges 
briefly mentioned before) a profile function in transverse space is 
introduced so to enter into a momentum charge that contributes to 
the BPS charge. These profile functions then enter into the 
supergravity solutions that are supposed to hold in the strong 
coupling regime and change their behaviour at short radius, 
differently for every different profile function. They are not 
singular at r=0 and the value of r where they all start resembling 
the usual b.h. solution outside the horizon is identified as a fuzzy 
``horizon'' of a fuzzball proposal for b.h.. It is unclear, however, 
if and how a trapped region could emerge for the incoherent 
superposition characterizing the b.h. macrostate.

\section{The role of decoherence}
If string microstates counted by the entropy of b.h. macrostates are 
not themselves black holes, it should happen that decoherence, 
intrinsic to any classical limit, should be critical in building b.h. 
characteristics as metric singularities and event horizons. It is 
not surprising that decoherence may have an important role in high 
excitation string physics due to the very large degeneracy of states 
in that regime. Indeed, even for $g \to 0$, i.e. tree 
diagrams, the non-trivial spectrum of emitted particles in the decay 
of any high mass excitation gives rise to a thermal distribution if 
an average over the very many states with the same mass is 
performed~\cite{16}.  

Even if effects of this kind may well be at work also for large 
couplings, decoherence should have much more subtle effects in order 
to generate b.h. physics from non black-holish microstates. Let me 
provide some speculative ideas on how a geometric picture could 
arise from a decoherence procedure in the pregeometric string 
approach. In this theory, indeed, even space and time are defined 
through the string; they are operators and not parameters that could 
be interpreted as coordinates of a space-time that may subtend a 
dynamical geometry. These are all concepts that may arise in a 
classical limit of the theory when quantum fluctuations may be 
neglected.  But even in this limit, the theory contains in principle 
not only the metric and possibly matter fields, but also an infinite 
number of higher tensor fields whose effect may possibly be ignored 
only in some conditions. The (infinite number of) equations that 
these (infinite number of) fields should satisfy, are given by the 
condition of no conformal anomaly ($\beta = 0$), and it is in the 
limit of small frequencies (in string length units) that only 
massless fields appear satisfying Einstein's equations~\cite{17}. 
But in presence of a horizon of a metric solution, the statement of 
low frequency is not relativistic invariant. Indeed, an arbitrary low 
frequency wave for a fixed external observer will be perceived by a 
free falling one with a blue shift which gets arbitrarily increased 
when approaching the horizon. This means that to have disregarded 
contributions with higher derivatives, or fields with higher 
tensorial character, would have been an unwarranted approximation. 
And even a small effect of those tensors could have avoided the 
metric condition that implied the singularity and the trapped region 
in the usual Einstein equation.  There could be many solutions 
involving different field configurations in which the metric and 
other tensor fields are classically entangled with relevant phases.  
And it could well happen that an incoherent superposition of these 
different background configurations could wash out the higher 
tensorial fields leaving a geometric description with, eventually, a 
b.h. metric with its singularity and its event horizon. This could 
be a hypothetical way in which non b.h. microstates could give rise 
to a b.h. macrostate.

In this case, the apparent contradiction between b.h. in classical 
general relativity and quantum coherence is solved in a conceptually 
simple way: it is the decoherence procedure, implied in any 
classical limit, that gives rise --- from a consistent quantum 
theory of gravitation as superstrings --- to a classical 
geometrical space-time description (general relativity) with 
eventual trapped regions, event horizon and b.h. and, of course, the 
loss of quantum coherence.

\section{High energy collisions in string theory and metric back reaction}
Let us now discuss high energy scattering. Superstrings provide a 
computational perturbative algorithm for S-matrix amplitudes that, 
if properly resumed, allows an explicit analysis of the continuation 
to the strong coupling (b.h.) regime. Therefore, as we shall discuss 
later, the consistent quantum theory may investigate situations in 
which, semiclassically, the process should be described by a b.h. 
formation and subsequent evaporation. Thus, hopefully, the analysis 
may throw light on how and why may happen that a coherent quantum 
state would not produce a b.h. even if the classical conditions to 
form it are met.

Much work has been done to study trans-Planckian collisions in a string approach [18,19].  I will recall methods and results that are consistently computable in the string regime and organized in an effective action form~\cite{20} to tackle their extention to a strong coupling regime where, semiclassically, b.h. formation and subsequent evaporation should be expected.
As already said, string (or actually superstring) theories contain a dimensional scale --- the string length $l_s$ --- and a dimensionless one g, the string coupling that generates the genus expansion.  Gravitational scales, as the Newton constant G or the Schwarzschild radius $R_S$ corresponding to an energy  E are given by
\begin{equation}
G=g^2 l^2_s/\h \qquad \qquad R_S=GE
\label{eqn1}
\end{equation}
For simplicity, eqs~(\ref{eqn1}) and other explicit expressions we shall give refer to the $d=4$ case even if the analysis we  recall has been done for arbitrary d non compactified dimensions.
The method used in refs.[18] is to consider a trans-Planckian regime defined by a small coupling-large energy 
\begin{equation}
g^2 << l,\, E l_s /\h >> 1 \label{eqn2}
\end{equation}
so that 
\begin{equation}
G E^2/\h=g^2 (E l_s/\h)^2 >1 \label{eqn3}
\end{equation}
In the genus expansion of string amplitudes all terms in which $g^2$ is enhanced by the large factor as in eq.~(\ref{eqn3}) have to be considered and resumed. Let us notice that in the large energy regime of eqs.~(\ref{eqn2}),~(\ref{eqn3})  $R_S/l_s = g^2 E l_s/\h$  can be smaller or larger than one and, as we shall see, physics will be different on the two sides of the inequality.
The computation of the collision amplitude in superstring theory in terms of the enery E and impact parameter b has been organized in powers of  $R_S^2/b^2$.  For b larger than both $R_S$ and $l_s$,  the two particle collision amplitude in the high energy regime as defined by eq.~(\ref{eqn2}) --- obtained by the just discussed all order resummation --- has an eikonal form, the eikonal  being a Hermitian operator (thus unitary S-matrix) in the Fock space of the two colliding strings.
Only for very large values of b --- where the amplitude is perturbative and dominated by the graviton pole --- the scattering is elastic, while for  $b < g E l_s^2/\h$  the two colliding gravitons are also excited to other superstring states in the scattering process. The eikonal is large and allows a classical trajectory interpretation through a saddle point in the Bessel transform to transfer momentum.  It reproduces the relation between deflection angle and impact parameter classically experienced by each particle in the gravitational field (Aichelburg-Sexl) created by the other one. With the extra fact that while deflecting, colliding particles may be excited (in a calculable way) to one of its string recurrences, implying an attenuation of the elastic amplitude (imaginary phase) that increases, together with the deflection angle, for decreasing b. 
In the $R_S < l_s$ case, b may decrease where string effects become relevant, giving rise to copious inelastic production~\cite{21} and thus a softening that implies an attenuation of the elastic amplitude and a reduced deflection angle. 
In the $R_S > l_s$ case, when b approaches $R_S$ new terms appear, as said before, in the form of powers of $R_S^2/b^2$ that look as classical corrections despite their quantum origin.  The first term has been computed in the string framework~\cite{15,22} and an effective action algorithm has been proposed for computing and resuming them all~\cite{20}. This may be interpreted as a metric and dilaton background generated by the process or, equivalently, a consistent quantum computation of back reaction on the metric, giving effects that become relevant when approaching situations in which a b.h formation is classically expected.  It could thus represent a way of understanding how and why a b.h. is avoided in a well defined quantum state as that under discussion. 
It is perhaps unfortunate that no further effort has been devoted in that direction. I have even a vague recall of a sense of frustration of the scientist to whom we dedicate these contributions, Ciafaloni and myself when --- many years ago --- some preliminary results could not be forced into the recognition of a horizon. Fact that brought us to give up, while today I would consider it as the expected sign to reveal novel quantum gravitational effects!
Furthermore, if this sort of back reaction is efficient in avoiding trapped regions in the well defined quantum state represented by the two colliding particles, it could perhaps continue to do so in arbitrary collapse situations. Let me also adventure that this possible effect of quantum back reactions on the metric may allow an interpretation of the recent Hawking suggestion~\cite{2} that the original classical solution, as the Schwarzschild metric in a gravitational collapse, may give an irrelevant contribution to the path integral for the actual gravitational process.

\section{Metric back reaction and possible avoidance of  blackholes}
The idea that that standard b.h. may not be the objects realized in nature even at the macroscopic level, has been recently explored within different contexts~\cite{23}. In particular, interesting suggestions have been borrowed from geometric acoustical models that can be studied experimentally and show a physics that is associated with classical and quantum fields in curved space-times~\cite{24}. 
Propagation of small disturbances in the flow of even simple fluids are known to behave equivalently to a linear (classical or quantum) field over an acoustic space-time endowed with an acoustic metric~\cite{25}. Depending on that endowed metric, acoustic b.h. -trapped region corresponding to a supersonic regime in fluid flow- may be created.  It has been however noted~\cite{26} that Hawking-like radiation does not necessarily imply the formation of a trapped region; it is sufficient that a sonic point conveniently develops in the asymptotic future. The radiation is then controlled by a temperature that contains both the Hawking one and the rate by which the sonic point is reached for $t \to \infty$.  This critical collapse result suggests an alternative scenario for a semiclassical collapse and evaporation of ``b.h.'' objects that --- very speculatively --- could be exported to semiclassical gravity. Its interpretation would imply that some quantum back-reaction on the geometry could prevent the surface of the collapsing star (or impinging matter) from actually crossing the Schwarzschild radius. At later stages, the evaporation process would become more efficient so to induce a chasing of the would be horizon by the surface of the star that could end with the complete evaporation and a flat space-time~\cite{26}.

\section{Conclusions and outlook}

I hope to have substantiated my (probably personal) point of view of why some coincidences between string state multiplicity and average decay spectra, on one hand, and b.h. entropy and evaporation  spectra, on the other, are far from having solved in a naive way  the apparent paradox of loss of quantum coherence in b.h. formation and evaporation (the information paradox). String microstates, in particular, aren't b.h. and well defined quantum states would not generate b.h. even if they would have been expected on classical grounds.
I have discussed two ways to resolve this apparent discrepancy, both of them accessible to further investigation in the string framework.
The first one starts from the fact that superstring theory is pregeometrical and even the concept of space-time is induced by the string through a classical limit. Thus space-time, geometry, event horizons, black holes and the loss of quantum coherence would all come with the same token i.e. the decoherence procedure implied in the classical limit that leads to general relativity. Thus no paradox: either bona fide quantum (as superstrings) or classical space-time with dynamical geometry and black holes but no a priori quantum coherence.
The other possibility is that the lack of b.h. formation in a quantum state, as two particle collision, may be due to well identified quantum contributions that give rise to apparently classical effects that act as quantum back reactions on the metric. Effects that could remain influential even in classical gravitational collapse processes thus avoiding metric singularities, trapped regions and event horizons. Without forming, therefore, even classical b.h. despite the fact that many external observational properties would not look very dissimilar. Thus no paradox because no real black holes: no trapped region or event horizon to spoil quantum coherence or information retrieval.

\section*{Recognition}
I had the chance to enjoy a lively and fruitful  collaboration with Gabriele for many years and on a variety of subjects.  Sharing --- as also reflected in this paper --- the joy of elaborating original physics, the frustration of unexpected obstacles and the persisting challenge of different viewpoints on possible developments.  I wish him to keep harvesting success, surrounded by friends and collaborators attracted by his scientific and human qualities. People of all origins and ages...... with me at the oldest end.

\printindex
\end{document}